\documentstyle[a4,12pt,epsf]{article}
\epsfverbosetrue
\newcommand{\figurebox}[2]{\fbox{\vbox to#2in{\hbox to #1in{\hfil} \vfil}}}
\newcommand{\err}[2]{{\footnotesize {$\;\begin{array}{@{}l@{}}
			  +\makebox[1.25em][r]{#1} \\[-0.6em]
			  -\makebox[1.25em][r]{#2}
			\end{array}$}}}
\newcommand{\widerr}[2]{{\footnotesize {$\,\begin{array}{@{}l@{}}
			  +\makebox[1.4em][r]{#1} \\[-0.6em]
			  -\makebox[1.4em][r]{#2}
			\end{array}$}}}
\newcommand{\beq}{\begin{equation}}
\newcommand{\eeq}{\end{equation}}
\newcommand{\dleft}{\stackrel{\leftarrow}{D}}
\newcommand{\dright}{\stackrel{\rightarrow}{D}}
\begin{document}

\begin{titlepage}

\begin{flushright}
Edinburgh Preprint: 92/506\\
Southampton Preprint: SHEP 91/92--15\\
Revised, 10 April, 1992
\end{flushright}
\vspace{5mm}
\begin{center}
{\Huge Quenched Hadrons using Wilson and $O(a)$-Improved Fermion
Actions at $\beta=6.2$}\\[15mm]
{\large\it UKQCD Collaboration}\\[3mm]

{\bf C.R.~Allton, H.~Duong, C.T.~Sachrajda}\\
Physics Department, The University, Southampton SO9~5NH, UK

{\bf R.M.~Baxter, S.P.~Booth, K.C.~Bowler, S.~Collins, D.S.~Henty,
R.D.~Kenway, C.~McNeile, B.J.~Pendleton, D.G.~Richards, J.N.~Simone,
A.D.~Simpson, B.E.~Wilkes}\\
Department of Physics, The University of Edinburgh, Edinburgh EH9~3JZ,
Scotland

{\bf A.~Hulsebos, A.C.~Irving, A.~McKerrell, C.~Michael, M.~Prisznyak,
P.W.~Stephenson}\\
DAMTP, University of Liverpool, Liverpool L69~3BX, UK

{\bf G.~Martinelli}\\
Dipartimento di Fisica, Universit\`a di Roma {\em La Sapienza},
00185 Roma,\\and INFN Sezione di Roma, Roma, Italy

\end{center}
\vspace{5mm}
\begin{abstract}

We present the first study of the light hadron spectrum and decay
constants for quenched QCD using an $O(a)$-improved
nearest-neighbour Wilson fermion action at $\beta=6.2$.  We
compare the results with those obtained using the standard Wilson
fermion action, on the same set of 18 gauge field configurations
of a $24^3\times 48$ lattice.  For pseudoscalar meson masses in
the range 330-800~MeV, we find no significant difference between
the results for the two actions.  The scales obtained from the
string tension and mesonic sector are consistent, but differ from
that derived from baryon masses.  The ratio of the pseudoscalar
decay constant to the vector meson mass is roughly independent of
quark mass as observed experimentally, and in approximate
agreement with the measured value.

\end{abstract}

\end{titlepage}

\paragraph{Introduction}\label{introduction}

The Wilson formulation of lattice fermions leads to matrix
elements which differ from their continuum counterparts by terms
of order the lattice spacing, $a$.  These corrections, which
cannot be calculated reliably in perturbation theory, represent a
substantial uncertainty in the determination of hadronic matrix
elements.  A tree-level improved theory has been
proposed~\cite{sheikholeslami,hmprs}, which in perturbation theory
eliminates all terms of order $(g_0^2)^na\log^na$~\cite{hmprs}.
Our objective, here, is to look for evidence of such improvement
in masses and decay constants of the light hadrons.  In this
letter, we summarise the results of a study in quenched QCD on a
set of 18 configurations of a $24^3\times 48$ lattice at
$\beta=6.2$, using both the standard Wilson fermion action:
\begin{equation}
S_F^W  =  a^4\sum _x \frac{1}{a}\Biggl\{\bar{q}(x)q(x)
             +\kappa\sum _\mu\Bigl[
            \bar{q}(x)(\gamma _\mu -r)U_\mu (x) q(x+\hat\mu ) -
            \bar{q}(x+\hat\mu )(\gamma _\mu + r)U^\dagger _\mu(x)
            q(x)\Bigr]\Biggr\}
\end{equation}
and the nearest-neighbour $O(a)$-improved, or ``clover'', fermion
action~\cite{sheikholeslami}:
\begin{equation}
S_F^C  = S_F^W - irg_0\kappa\frac{a}{2}a^4\sum_{x,\mu,\nu}\bar{q}(x)
         F_{\mu\nu}(x)\sigma_{\mu\nu}q(x).
\end{equation}
$F_{\mu\nu}$ is a lattice definition of the field strength tensor,
which we take to be the sum of the four untraced plaquettes in the
$\mu\nu$ plane with a corner at the point
$x$~\cite{sheikholeslami}:
\begin{equation}
\label{eq:clover_term}
F_{\mu\nu}(x) = \frac{1}{4} \sum_{\Box=1}^{4} \frac{1}{i2g_0a^2}
                \biggl[U_{\Box\mu\nu}(x) - U_{\Box\mu\nu}^\dagger(x)\biggr].
\end{equation}
Full details of the simulations will be presented
elsewhere~\cite{hadrons_paper}.

\paragraph{Computational Details}\label{simulation}

The gauge field configurations and quark propagators were obtained
using the 64 i860-node Meiko Computing Surface at Edinburgh.  The
full machine has 1~GByte of internal memory and, for these
programs, sustains around 1.5~Gflops.

We update the SU(3) gauge fields with a cycle of 1 three-subgroup
Cabibbo-Marinari heat-bath sweep followed by 5 over-relaxed
sweeps.  The results presented here are based on an analysis for
each action of the same set of 18 configurations, starting at
configuration 16800 and separated by 2400 sweeps.  Propagators are
calculated for $r=1$ at $\kappa = $ 0.1510, 0.1520, 0.1523, 0.1526
and 0.1529 for the Wilson action, and at $\kappa = $ 0.14144,
0.14226, 0.14244, 0.14262 and 0.14280 for the clover action; the
latter values were chosen to match roughly the pion masses
computed in the Wilson case.  We use an over-relaxed minimal
residual algorithm with red-black preconditioning for propagator
calculations, with point sources and sinks.

\paragraph{String Tension}

It is useful to have an estimate of the scale from a quantity
which is independent of the fermion action.  For this reason, we
present results for the string tension obtained from generalised
$R\times T$ Wilson loops, using a variational scheme comprising
$40$ and $28$ recursive blockings~\cite{albanese} for the spatial
paths.  We obtain the potential by extrapolating the $3:2$ and
$4:3$ $T$-ratio results using an estimate of the excited state
contribution.

We find an acceptable fit to the static potential for $2 \leq R/a
\leq 12$ with the conventional lattice Coulomb plus linear term:
\begin{equation}
V(R) = C - {E\over R} + K R
\end{equation}
and obtain $E= 0.274(6)$ and $Ka^2=0.0259(9)$, where the quoted
errors are purely statistical.  Using this value for the string
tension to set the scale, by requiring $\sqrt K=0.44$ GeV, yields
$a^{-1} = 2.73(5)$ GeV.  These results are in agreement with a
similar analysis on $20^{4}$ lattices at the same $\beta
$-value~\cite{perandmich,michandtep}.  We find from a comparison
with lower $\beta $ results, that asymptotic scaling of the string
tension does not hold at $\beta =6.2.$

\paragraph{Hadron Spectrum}

We identify the timeslice interval 12 to 16 as a common plateau
region in the effective mass plots.  We obtain the amplitudes and
masses of mesons (baryons) by correlated least-$\chi^2$ fits of a
single cosh (exponential) function simultaneously to the
appropriate forward and backward propagators.  The fits all have
$\chi^2$/dof in the range 0.3 to 3.  The quoted errors are 68\%
confidence level bounds estimated by a bootstrap procedure.  The
details of our statistical analysis will be presented
elsewhere~\cite{hadrons_paper}.

\begin{figure}[htbp]
\begin{center}
\leavevmode
\epsfysize=300pt
  \epsfbox[20 30 620 600]{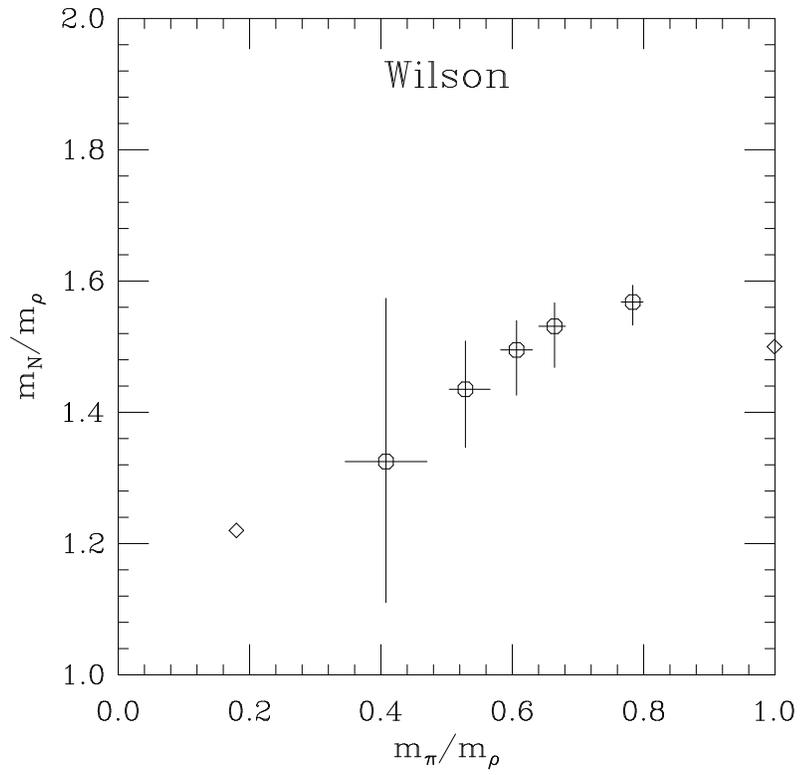}
\end{center}
\begin{center}
\leavevmode
\epsfysize=300pt
  \epsfbox[20 30 620 600]{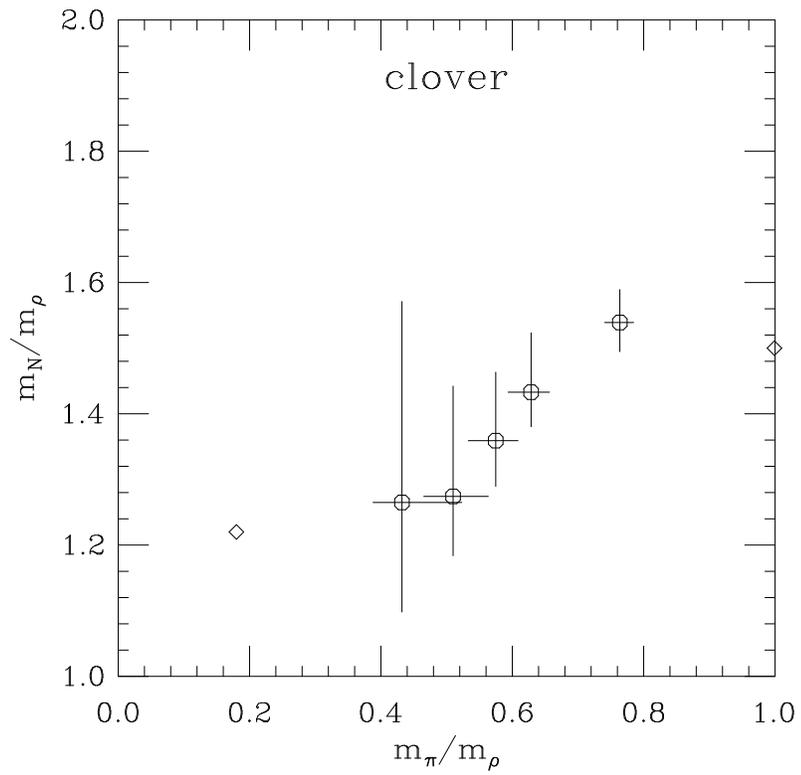}
\end{center}
\caption{Edinburgh plots.}
\label{edin-plots}
\end{figure}

The pion, $\rho$, nucleon and $\Delta$ mass estimates in lattice units
for the two actions are given in table~\ref{tab:masses}.
\begin{table}
\begin{center}
\begin{tabular}{|c|l|l|l|l|}
\hline
\multicolumn{5}{|c|}{Wilson}\\\hline
$\kappa$
& \multicolumn{1}{|c|}{$am_\pi$}
& \multicolumn{1}{|c|}{$am_\rho$}
& \multicolumn{1}{|c|}{$am_N$}
& \multicolumn{1}{|c|}{$am_\Delta$}
\\
\hline
0.1510  & 0.295\err{6}{3}  & 0.377\err{11}{5}  & 0.591\err{11}{6}  &
0.647\err{19}{10}  \\
0.1520  & 0.221\err{9}{3}  & 0.332\err{15}{4}  & 0.509\err{15}{7}  &
0.582\err{19}{9}  \\
0.1523  & 0.195\err{9}{3}  & 0.321\err{16}{6}  & 0.480\err{16}{9}  &
0.560\err{23}{11}  \\
0.1526  & 0.164\err{11}{4} & 0.310\err{19}{11} & 0.445\err{16}{13} &
0.538\err{36}{18} \\
0.1529  & 0.122\err{9}{11} & 0.298\err{39}{34} & 0.395\err{48}{42} &
0.533\widerr{116}{40}\\
\hline
0.15329\err{7}{4}
	& 0.0              & 0.278\err{25}{9} & 0.393\err{20}{16} & 0.497\err{31}{16}
\\
\hline\hline
\multicolumn{5}{|c|}{clover}\\\hline
$\kappa$
& \multicolumn{1}{|c|}{$am_\pi$}
& \multicolumn{1}{|c|}{$am_\rho$}
& \multicolumn{1}{|c|}{$am_N$}
& \multicolumn{1}{|c|}{$am_\Delta$}
\\
\hline
0.14144 & 0.302\err{6}{4}  & 0.395\err{13}{9}  & 0.608\err{15}{8}  &
0.678\err{15}{12} \\
0.14226 & 0.217\err{8}{6}  & 0.345\err{18}{10} & 0.495\err{30}{8}  &
0.598\err{25}{15} \\
0.14244 & 0.194\err{9}{6}  & 0.338\err{23}{13} & 0.460\err{34}{11} &
0.577\err{25}{22} \\
0.14262 & 0.168\err{11}{6} & 0.329\err{30}{23} & 0.419\err{45}{17} &
0.565\err{33}{33} \\
0.14280 & 0.135\err{11}{6} & 0.313\err{35}{44} & 0.396\err{51}{38} &
0.586\err{88}{62}
\\
\hline
0.14314\err{6}{4}
        & 0.0              & 0.293\err{26}{21} & 0.376\err{42}{18} &
0.513\err{40}{31} \\
\hline
\end{tabular}
%
%
\caption{{\it Hadron masses in lattice units. The last row for each
action contains the values obtained by linear extrapolation to
$am_\pi=0$.}\label{tab:masses}
}
\end{center}
\end{table}
Edinburgh plots for the two actions are given in
fig.~\ref{edin-plots}.  The plots are broadly consistent, showing
a trend towards the physical value for $m_N/m_\rho$ with
decreasing pion mass.

\begin{figure}[htbp]
\begin{center}
\leavevmode
\epsfysize=300pt
  \epsfbox[20 30 620 600]{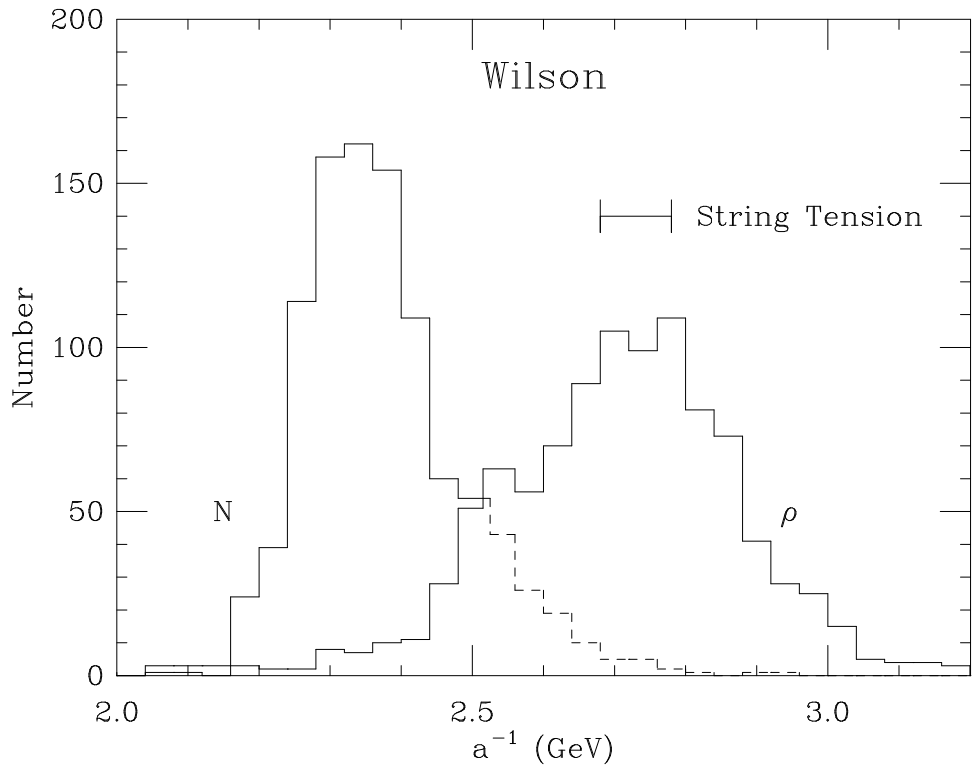}
\end{center}
\begin{center}
\leavevmode
\epsfysize=300pt
  \epsfbox[20 30 620 600]{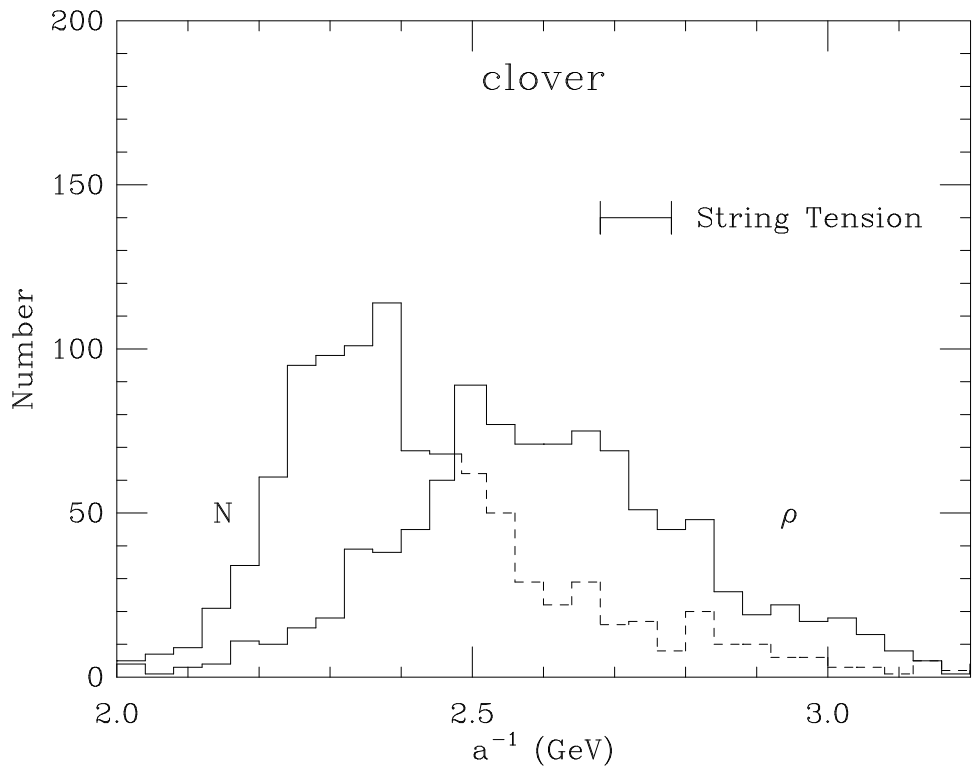} \end{center}
\caption{Distributions of the scales determined from $m_\rho$ and
$m_N$ for 1000 bootstrap samples compared with the scale from the
string tension.} \label{scales}
\end{figure}

We now consider chiral extrapolation of the hadron masses in order
to compute the lattice scale for the two actions.  We perform
correlated linear fits to the data at all five $\kappa$ values.
{}From the bootstrap analysis, we find that there are strong
correlations between the pion masses at different $\kappa$ values.
The behaviour of $m_\pi^2$ as a function of $1/2\kappa$ for both
actions is completely consistent with PCAC behaviour throughout
the quark mass range used.  From the chiral extrapolation, we
obtain
\begin{equation}
\kappa_c = 0.15329\mbox{\err{7}{4}}\hspace{0.3cm} {\rm (Wilson)},\hspace{1cm}
\kappa_c = 0.14314\mbox{\err{6}{4}}\hspace{0.3cm} {\rm (clover)}.
\end{equation}
The lattice scales obtained from correlated linear extrapolations
of the other hadron masses to the chiral limit are presented in
table~\ref{tab:scales}.
\begin{table}
\centering
\begin{tabular}{|c|l|l|}\hline
                                  & \multicolumn{2}{c|}{$a^{-1}$(GeV)} \\\hline
method                            & Wilson          &  clover     \\\hline
string tension                    & 2.73\err{5}{5}  &  2.73\err{5}{5}  \\
$m_\rho$                          & 2.77\err{9}{23} &  2.62\err{11}{21}\\
$m_N$                             & 2.39\err{10}{12}&  2.49\err{12}{24}\\
$m_\Delta$                        & 2.48\err{8}{15} &  2.40\err{15}{17}\\\hline
\end{tabular}
\caption{{\it Scales determined from different physical quantities.}}
\label{tab:scales}
\end{table}
For both actions, the scales derived from $m_\rho$ agree well with
the scale from the string tension, whereas the scales from the
baryon masses, whilst consistent with one another, are too low, in
common with previous studies using the Wilson action.  This
discrepancy between the baryon and string tension scales is
clearly seen in the estimate of the probability distributions for
the scales obtained from the bootstrap samples, shown in
figure~\ref{scales}.

\begin{figure}[htbp]
\begin{center}
\leavevmode
\epsfysize=300pt
  \epsfbox[20 30 620 600]{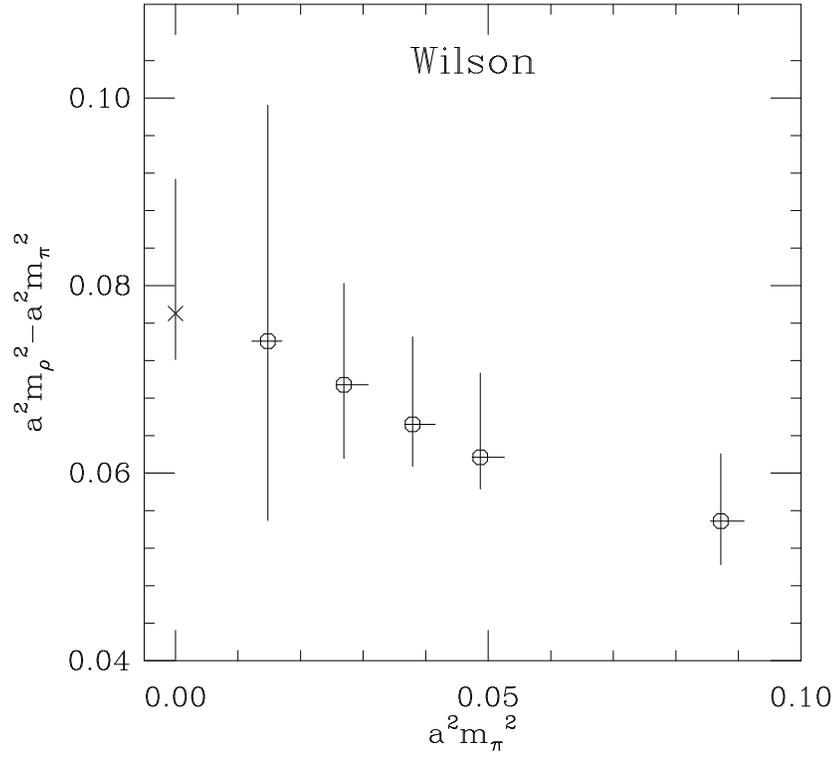}
\end{center}
\begin{center}
\leavevmode
\epsfysize=300pt
  \epsfbox[20 30 620 600]{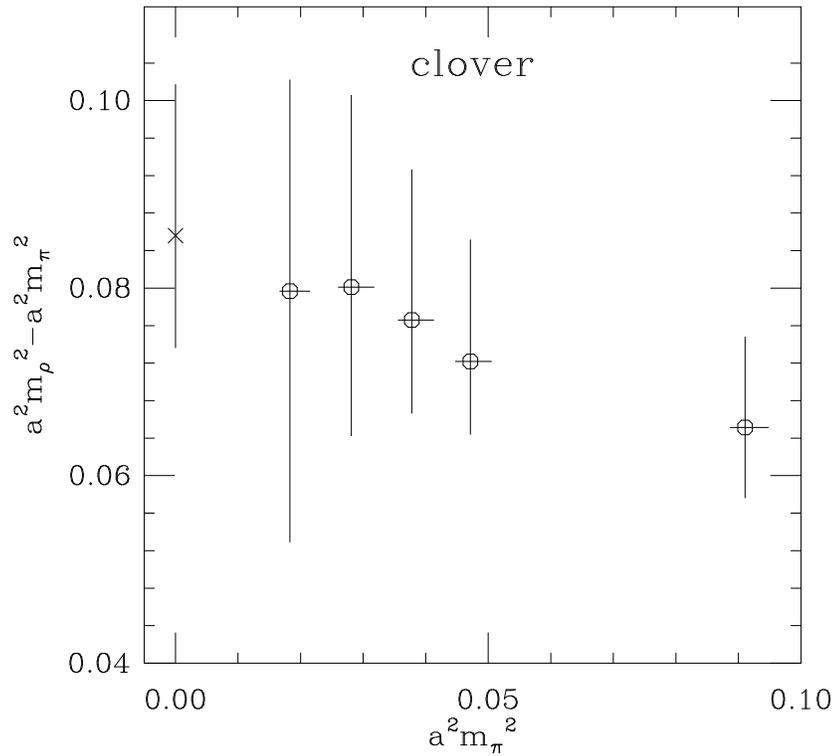}
\end{center}
\caption{$a^2m_\rho^2-a^2m_\pi^2$ versus $a^2m_\pi^2$, in lattice units; the
left-most point in each plot is obtained from the chiral
extrapolation of the individual masses.} \label{vec-pseudo}
\end{figure}

In an attempt to highlight any differences arising from use of the
clover action, we present estimates for the vector-pseudoscalar
meson mass splitting, which in QCD-inspired quark models arises
from spin interactions and therefore may be corrected at $O(a)$ by
the spin term in the clover action.  Experimentally, for both
light-light and heavy-light mesons, $m_V^2-m_P^2$ is very nearly
independent of quark mass: 0.57~GeV$^2$ ($\rho-\pi$), 0.55~GeV$^2$
($K^\ast-K$, $D^\ast-D$, $B^\ast-B$).  Taking the scale from
the string tension, the range of pion masses for which we have
data is 330~MeV to 800~MeV.  Thus, it is of interest to ask
whether our data for $m_\rho^2-m_\pi^2$, with equal-mass quarks,
has similar behaviour.  In figure~\ref{vec-pseudo}, we plot the
quantity $(am_\rho)^2-(am_\pi)^2$, calculated directly from the
bootstrap masses, versus $(am_\pi)^2$ for both actions.  The
Wilson data is consistent with previous work~\cite{hyperfine}
which indicated a negative slope, inconsistent with experiment at
large quark mass (e.g.~for the charmonium system $m_V^2-m_P^2 =
0.72$~GeV$^2$).  Apart from a constant vertical shift, the best
estimates from our clover data at the four highest quark masses
agree with the Wilson points; the larger statistical errors in the
clover data leave open the possibility of a reduced, or zero
dependence on quark mass.  Using the string tension scale, the
experimental value corresponds to 0.075 in lattice units,
consistent with the three lightest Wilson points and all the
clover points.

\paragraph{Meson Decay Constants}

The pion decay constant is defined through the matrix element of
the axial current:
\begin{equation}
\langle0|\bar q(0)\gamma _\mu\gamma_5q(0)|\pi (p)\rangle=f_\pi p_\mu
\end{equation}
and our normalisation is such that the physical value is 132 MeV.  In
table~\ref{tab:fpi}, we present the values obtained for $f_\pi$ and
for the dimensionless ratio $f_\pi/m_\rho$, using the Wilson
action with the local axial current, and the clover action with the
``improved'' axial current:
\begin{equation}
\bar q(x)(1+\frac{ra}{2}\gamma\cdot\dleft )\gamma_\mu\gamma_5
(1-\frac{ra}{2}\gamma\cdot\dright )q(x).
\end{equation}
Our lattice results for $f_\pi/m_\rho$ vary only slowly with quark
mass, in agreement with the experimental observation that
$f_\pi/m_\rho$ (0.17) is approximately the same as
$f_K/m_{K^\ast}$ (0.18).
\begin{table}
\centering
\begin{tabular}{|c|c|c||c|c|c|}\hline
\multicolumn{3}{|c||}{Wilson} & \multicolumn{3}{c|}{clover}\\\hline
$\kappa$ & $(af_\pi)/Z_A^W$ & $(f_\pi/m_\rho )/Z_A^W$ &
$\kappa$ & $(af_\pi)/Z_A^C$ & $(f_\pi/m_\rho )/Z_A^C$ \\ \hline
0.1510   & 0.081\err{3}{2}  & 0.22\err{1}{1} & 0.14144 & 0.060\err{5}{3} &
0.15\err{1}{1} \\
0.1520   & 0.069\err{4}{5}  & 0.21\err{1}{2} & 0.14226 & 0.048\err{7}{3} &
0.14\err{2}{1} \\
0.1523   & 0.066\err{4}{8}  & 0.21\err{1}{3} & 0.14244 & 0.046\err{8}{5} &
0.14\err{2}{2} \\
0.1526   & 0.065\err{5}{11} & 0.21\err{2}{4} & 0.14262 & 0.046\err{8}{6} &
0.14\err{3}{2} \\
0.1529   & 0.067\err{6}{13} & 0.23\err{3}{5} & 0.14280 & 0.046\err{11}{9}&
0.15\err{5}{3} \\ \hline
0.15329  & 0.057\err{6}{8}  & 0.21\err{2}{3} & 0.14314 & 0.037\err{10}{7}&
0.13\err{3}{3} \\ \hline
\end{tabular}
\caption{{\it Values of the pion decay constant, in lattice units, and the
ratio $f_\pi/m_\rho$.
The last row contains values obtained by a linear extrapolation to the
chiral limit.}}
\label{tab:fpi}
\end{table}

\begin{figure}[htbp]
\begin{center}
\leavevmode
\epsfysize=300pt
  \epsfbox[20 30 620 600]{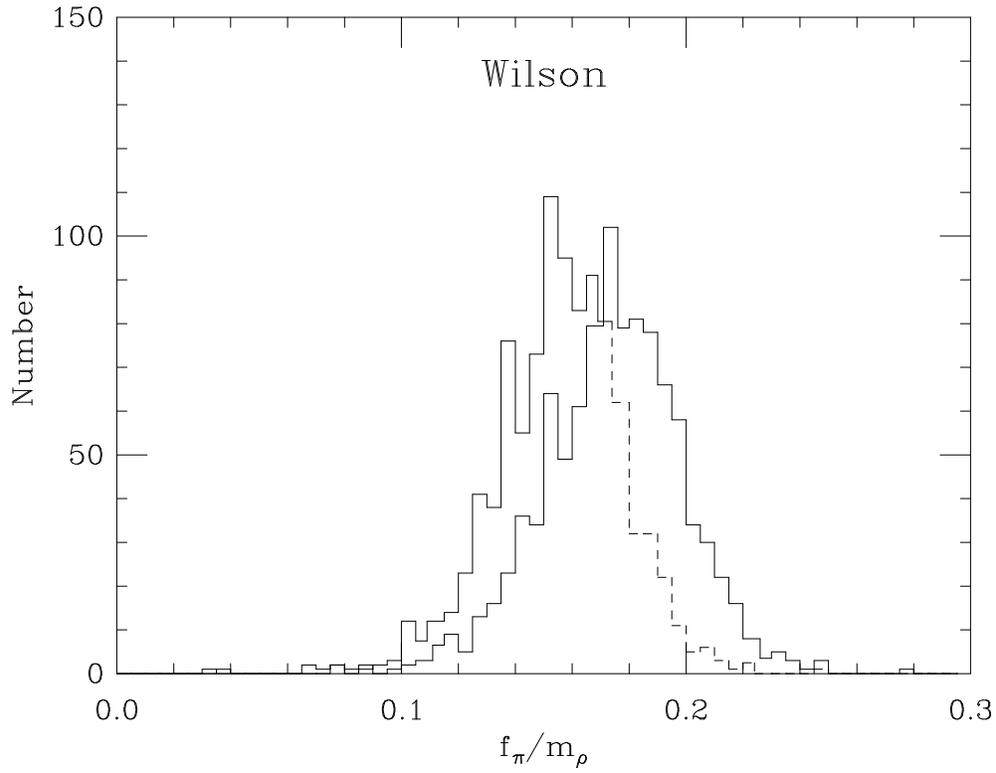}
\end{center}
\begin{center}
\leavevmode
\epsfysize=300pt
  \epsfbox[20 30 620 600]{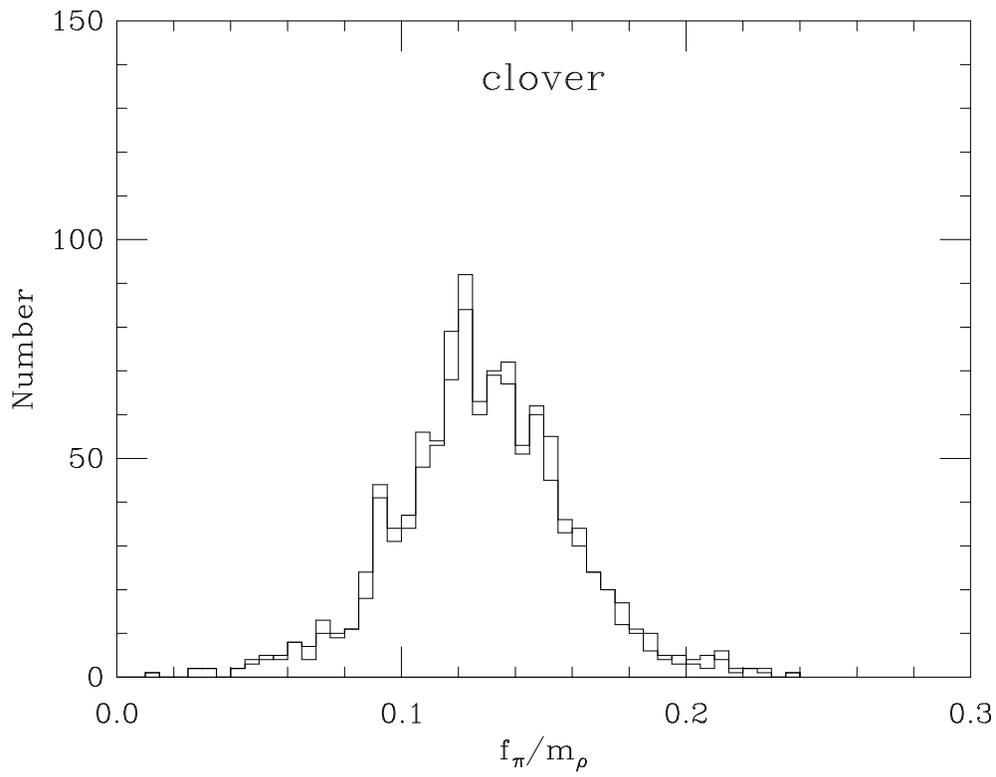}
\end{center}
\caption{Distributions of $f_\pi/m_\rho$ corresponding to the two
choices for $Z_A^W$ and $Z_A^C$ described in the text, for 1000
bootstrap samples.}
\label{fpi/mrho}
\end{figure}

In order to determine the physical values, the lattice results
given in table~\ref{tab:fpi} need to be multiplied by the
appropriate renormalisation constant, $Z_A^W$ or $Z_A^C$.
$Z_A^W\simeq 1-0.132g^2$ in perturbation theory.  If we use the
bare coupling constant as the expansion parameter, then
$Z_A^W\simeq 0.87$ and we find, for the Wilson action,
$f_\pi/m_\rho\simeq 0.18$\err{2}{3}.  If instead we use the
``effective coupling'' proposed in reference~\cite{lm}, $g^2_{\rm
eff} \simeq 1.75 g^2_0$, then $Z_A^W\simeq 0.78$ and we obtain
$f_\pi/m_\rho\simeq 0.16$\err{2}{3}.  The perturbative
estimate~\cite{pittori} of the renormalisation constant $Z_A^C$ is
close to 1 ($Z_A^C\simeq1-0.02g^2$); using the bare coupling leads
to $Z_A^C\simeq 0.98$, whereas the effective coupling gives
$Z_A^C\simeq 0.97$, yielding, for the clover action,
$f_\pi/m_\rho\simeq 0.12$\err{3}{2} in both cases.  We note that
the uncertainty due to the choice of the perturbative expansion
parameter is about 10\% with the Wilson action and only about 1\%
with the clover action.  Folding this in, the overall uncertainty
in the determination of $f_\pi/m_\rho$ may be gauged from the
bootstrap distributions in figure~\ref{fpi/mrho}.  It is apparent
that, within errors, the two actions give consistent
results.

In table~\ref{tab:frho}, we present values of the $\rho$ decay constant,
defined by the relation
\begin{equation}
\langle0|\bar q(0)\gamma _\mu q(0)|\rho\rangle=\frac{m_\rho ^2}{f_\rho}
\epsilon_\mu
\label{eq:frho}
\end{equation}
where $\epsilon_\mu$ is the $\rho$ polarisation vector.
\begin{table}
\centering
\begin{tabular}{|c|c||c|c|}\hline
\multicolumn{2}{|c||}{Wilson} & \multicolumn{2}{c|}{clover} \\ \hline
$\kappa$& $1/(Z_V^Wf_\rho)$ & $\kappa$ & $1/(Z_V^Cf_\rho)$ \\ \hline
0.1510  & 0.402\err{11}{13} & 0.14144 & 0.333\err{13}{19} \\
0.1520  & 0.431\err{10}{10} & 0.14226 & 0.359\err{15}{12} \\
0.1523  & 0.438\err{10}{8}  & 0.14244 & 0.364\err{16}{12} \\
0.1526  & 0.446\err{10}{10} & 0.14262 & 0.365\err{16}{12} \\
0.1529  & 0.465\err{22}{8}  & 0.14280 & 0.362\err{20}{14} \\ \hline
0.15329 & 0.470\err{18}{9}  & 0.14314 & 0.377\err{25}{12} \\ \hline
\end{tabular}
\caption{{\it Values of the $\rho$ decay constant.  The last row contains
values obtained by a linear extrapolation to the chiral limit.}}
\label{tab:frho}
\end{table}
For the Wilson action we use the local current and for the clover
action we use the improved local current.  Here, the
renormalisation constant for the Wilson action is given in
perturbation theory by $Z_V^W\simeq1-0.174g^2$.  Using the bare
(effective) coupling constant, $Z_V^W\simeq 0.83\ (0.71)$,
which leads to $1/f_\rho\simeq$~0.39\err{2}{1}
(0.33\err{1}{1}).  For the clover action, $Z_V^C\simeq 0.90\
(0.83)$~\cite{pittori} when the bare (effective) coupling
constant is used, which gives $1/f_\rho\simeq$~0.34\err{2}{1}
(0.31\err{2}{1}).  The physical value of $1/f_\rho$ is 0.28(1).
The uncertainty in the renormalisation constants makes it
difficult to draw firm conclusions about improvement from
$1/f_\rho$ computed with local currents.  The perturbative
uncertainty would be removed entirely by use of the conserved
vector current.  However, since the matrix element in
equation~(\ref{eq:frho}) is non-forward, the use of the conserved
current would not eliminate the $O(a)$ corrections in the Wilson
case.

\paragraph{Conclusions}

We have presented the first study of the light hadron spectrum and
decay constants for quenched QCD using an $O(a)$-improved
nearest-neighbour Wilson fermion action at $\beta=6.2$.  Having
compared the results with those obtained using the standard Wilson
fermion action, on the same set of 18 gauge field configurations
of a $24^3\times 48$ lattice, we see no statistically significant
difference between the best estimates for quantities derived from
2-point functions, for pseudoscalar meson masses in the range
330-800~MeV.  The clover data is typically noisier than the Wilson
data.  Nevertheless, there may yet be advantages from using the
clover action for 3-point functions, where improved chiral
behaviour for the $B$ parameter of $K$ decay has been
noted~\cite{Rome850}.  We find that at $\beta=6.2$ the scales
obtained from the meson sector are consistent with that from the
string tension, but there is clear evidence of an inconsistent
scale from the baryon sector.  $f_P/m_V$ is roughly independent of
quark mass as observed experimentally, and the numerical value
is broadly consistent with the measured value.

\paragraph{Acknowledgements}

This research is supported by the UK Science and Engineering
Research Council under grant GR/G~32779, by the University of
Edinburgh and by Meiko Limited.  We are grateful to Edinburgh
University Computing Service for use of its DAP 608 for some of
the analysis and, in particular, to Mike Brown for his tireless
efforts in maintaining service on the Meiko i860 Computing
Surface. CTS acknowledges the support of SERC through the award of
a Senior Fellowship.

\end{document}